\documentclass[reprint,superscriptaddress,amsmath,amssymb,aps,pra]{revtex4-2}

\usepackage{graphicx}
\usepackage{dcolumn}
\usepackage{bm}
\usepackage{physics}
\usepackage{amssymb}
\usepackage{textcomp}
\usepackage{amsthm}
\usepackage{mathtools}
\usepackage[utf8]{inputenc}
\usepackage[czech,polish,english]{babel}
\usepackage{dsfont}
\usepackage[shortlabels]{enumitem}
\usepackage[T1]{fontenc}
\usepackage{xcolor}
\usepackage{orcidlink}
\usepackage{hyperref}
\hypersetup{
    colorlinks,
    citecolor=blue,
    filecolor=black,
    linkcolor=blue,
    urlcolor=blue
}

\newcommand{\der}{\mathrm{d}}
\renewcommand{\Re}{\mathrm{Re}}
\renewcommand{\Im}{\mathrm{Im}}

\begin{document}

\title{Detection of quantum entanglement across the event horizon}

\author{Patryk Michalski\orcidlink{0009-0009-0305-7356}}
\email{pmichalski@cft.edu.pl}
\affiliation{Institute of Theoretical Physics, University of Warsaw, Pasteura 5, 02-093 Warsaw, Poland}
\affiliation{Center for Theoretical Physics, Polish Academy of Sciences, al. Lotnik\'{o}w 32/46, 02-668 Warsaw, Poland}

\author{Andrzej Dragan\orcidlink{0000-0002-5254-710X}}
\email{dragan@fuw.edu.pl}
\affiliation{Institute of Theoretical Physics, University of Warsaw, Pasteura 5, 02-093 Warsaw, Poland}
\affiliation{Centre for Quantum Technologies, National University of Singapore, 3 Science Drive 2, 117543 Singapore, Singapore}

\date{\today}

\begin{abstract}
We investigate the problem of distinguishing between separable and entangled states of two quantum wave packets, one of which falls into a black hole. Intuitively, one might expect the two scenarios to be indistinguishable, since the information carried by one wave packet is hidden beyond the event horizon. We show, however, that fundamental limitations on the localizability of quantum states can render the two scenarios, in principle, distinguishable. Employing tools from quantum state discrimination theory, we analyze a concrete realization and discuss the configurations that maximize the probability of successfully distinguishing between the two cases.
\end{abstract}

\maketitle

\section{Introduction}

One of the most striking predictions of classical general relativity is the existence of event horizons, acting as one-way causal boundaries beyond which no signal or influence can propagate to an external observer~\cite{Wald1984}. The presence of such a region of no escape is among the defining characteristics of a black hole~\cite{Curiel2019}. However, since the discovery of Hawking radiation~\cite{Hawking1974,Hawking1975} and related phenomena~\cite{Wald1994}, it has become clear that classical intuitions about black hole horizons are profoundly challenged once quantum effects are taken into account. In the absence of a fully satisfactory and consistent theory of quantum gravity, the problem can be approached within the framework of quantum field theory in curved spacetime, which describes the dynamics of quantum fields on a fixed classical gravitational background~\cite{Birrell1982}. Although inherently approximate, this framework is believed to offer valuable insights into what might be expected from a complete theory.

In this work, we employ the equivalence principle to examine a seemingly simple yet conceptually rich question that connects the notions of quantum entanglement, separability, and the presence of an event horizon. Previous analyses of entanglement in noninertial settings typically assumed observers on both sides of the horizon, and initially focused on global field modes~\cite{Alsing2003,Fuentes2005,Alsing2006,Martin-Martinez2009,Martin-Martinez2010,Martin-Martinez_Leon2010}. Subsequent studies identified interpretational difficulties associated with the single-mode approximation introduced in the pioneering results~\cite{Bruschi2010}, as well as obstacles in modeling realistic experimental configurations using global modes~\cite{Dragan_Doukas2013}. These concerns motivated the development of alternative approaches based on localized systems, including moving cavities~\cite{Downes2011,Bruschi2012,Friis2012}, point-like detectors~\cite{Lin2008,Lin_Chung2008,Lin2009,Doukas2010} and localized wave packets~\cite{Dragan2013,Richter2017}. The latter method was further generalized to accommodate wave packets with noncompact spatial support~\cite{Doukas2013}. Building on this framework, we revisit the question of entanglement across an event horizon from a fundamentally different perspective.

\begin{figure}[ht]
    \centering
    \includegraphics[width=\linewidth]{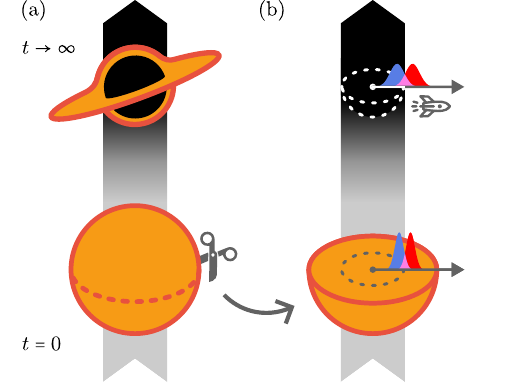}
    \caption{(a) A star undergoes gravitational collapse, forming a black hole in the asymptotic future. The dashed cut line indicates the spatial cross-section shown in the right panel. (b) During the collapse, at the location corresponding to the Schwarzschild radius (marked by the dashed circle), a quantum state is prepared using two orthogonal wave packets. In the asymptotic future, an observer Rob aboard a rocket measures the state of the field outside the event horizon.}
    \label{fig:1}
\end{figure}

Consider a gravitationally collapsing star in which, at the location corresponding to the Schwarzschild radius, a quantum field is prepared using two orthogonal wave packets (see Fig.~\ref{fig:1}). As the collapse proceeds, a black hole forms, causing the dominant part of one wave packet to fall behind the event horizon while the other predominantly escapes. Suppose a stationary observer, Rob, aboard a rocket measures the state of the field outside the horizon. A natural question then arises: can Rob determine whether the two wave packets were initially entangled or in a separable state?

At first sight, one might expect the two scenarios to be indistinguishable, since the information associated with one of the wave packets is irretrievably hidden beyond the event horizon. However, fundamental constraints on the localizability of quantum wave packets indicate that this expectation is not strictly valid. In particular, the Paley--Wiener theorem~\cite{Paley_Wiener} implies that any wave packet constructed exclusively from positive-frequency components cannot have compact support in position space and must exhibit a slower-than-exponential decay. Furthermore, the Reeh--Schlieder theorem~\cite{Reeh1961} demonstrates that the vacuum is cyclic for local field algebras, implying that no state of a quantum field can be perfectly localized within a bounded region \cite{Halvorson2001}. In addition, Hegerfeldt’s theorem~\cite{Hegerfeldt1974} imposes additional restrictions on the localization properties of states evolving under a positive-definite Hamiltonian. Consequently, one cannot exclude the possibility that the tails of a wave packet extend across the event horizon. Perhaps counterintuitively, we find that, once these constraints are taken into account, the entangled and separable cases may become, in principle, distinguishable.

The paper is organized as follows. In Sec.~\ref{sec:setup}, we introduce the framework used to model the two scenarios under consideration. Section~\ref{sec:state_discrimination} presents a concrete realization and quantifies the observable differences between the scenarios using tools from quantum state discrimination theory. Concluding remarks are given in Sec.~\ref{sec:conclusions}.

\section{Setup}\label{sec:setup}

For simplicity, we assume that the considered gravitational collapse produces a spherically symmetric black hole described by the Schwarzschild metric. The quantum state of the wave packets is assumed to be prepared before the collapse, when the space-time in the relevant region can be approximated as nearly flat (for example because the collapsing matter is sufficiently dilute). After the collapse, the wave packets propagate in the Schwarzschild geometry describing the exterior region of the black hole. Exploiting the symmetry of the spacetime, our analysis is restricted to the radial direction.

By the equivalence principle, the radial part of the Schwarzschild geometry in the immediate vicinity of the event horizon can be well approximated by the Rindler metric~\cite{Fuentes2005,Martin-Martinez2010}. In this picture, the freely falling wave packets correspond to states prepared by inertial observers, Alice and Bob, whereas a stationary observer located arbitrarily close to the Schwarzschild horizon is effectively modeled by an accelerated observer, Rob, positioned arbitrarily close to the Rindler horizon. As a further simplification that should still retain the qualitative features of the effect, we neglect the evolution of the wave packets during the collapse and treat their profiles at the moment of Rob’s measurement as given.

The quantum field under consideration is a real, massless scalar field $\hat{\phi}$ satisfying the Klein--Gordon equation $\square\hat{\phi} = 0$ in a two-dimensional spacetime. On the complex vector space of solutions to this equation, the (indefinite) Klein--Gordon inner product is defined as
\begin{equation}
    (\phi_1, \phi_2) = i \int_\Sigma \phi_1^* \overleftrightarrow{\partial_a} \phi_2\, n^a\, \mathrm{d}\Sigma,
\end{equation}
where $n^a$ denotes a future-directed normal vector to a spacelike hypersurface $\Sigma$, and $\mathrm{d}\Sigma$ is the corresponding volume element. All wave packets considered below are taken to be unit-normalized with respect to this inner product.

The Klein--Gordon equation can be solved in Minkowski coordinates $(ct,x)$, which are convenient for describing inertial observers. The positive-frequency mode solutions with respect to the Killing vector $\partial_t$ take the form
\begin{equation}
    u_k \equiv \frac{1}{\sqrt{4\pi c |k|}} \exp[i(kx - |k|ct)].
\end{equation}
Suppose that Alice and Bob can each prepare one of two orthogonal, possibly unlocalized, wave packets $\phi_\mathrm{A}(x,t)$ and $\phi_\mathrm{B}(x,t)$, respectively. We assume that both wave packets admit decompositions involving only positive-frequency Minkowski modes. Under this assumption, the operators $\hat{a} = (\phi_\mathrm{A},\hat{\phi})$ and $\hat{b} = (\phi_\mathrm{B},\hat{\phi})$ satisfy the canonical commutation relations and annihilate the Minkowski vacuum. This identifies $\hat{a}^{(\dag)}$ and $\hat{b}^{(\dag)}$ as the annihilation (creation) operators associated with the respective wave packets.

\begin{figure}[ht]
    \centering
    \includegraphics[width=\linewidth]{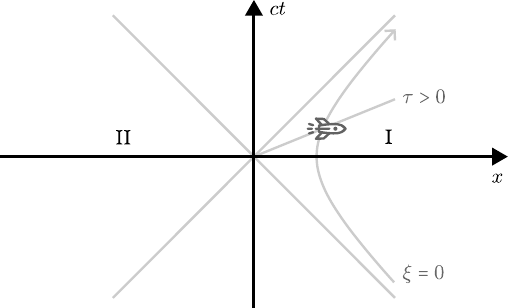}
    \caption{Minkowski diagram with Rindler regions~I and II. Region~I is covered with conformal Rindler coordinates $(c\tau, \xi)$. The world line $\xi = 0$ corresponds to a uniformly accelerated observer Rob.}
    \label{fig:rindler_chart}
\end{figure}

To describe Rob’s motion, we introduce conformal Rindler coordinates $(c\tau, \xi)$ parametrized by $a$:
\begin{equation}
    ct = \frac{c^2}{a} e^{a\xi/c^2} \sinh \frac{a \tau}{c}, \quad x = \frac{c^2}{a} e^{a\xi/c^2} \cosh \frac{a \tau}{c},
\end{equation}
which covers region~I ($x > c|t|$), as depicted in Fig.~\ref{fig:rindler_chart}. Rob travels along the world line $\xi = 0$, so that $a$ corresponds to his proper acceleration and $\tau$ to his proper time. In these coordinates, the region~I positive-frequency solutions of the Klein--Gordon equation with respect to the Killing vector $\partial_\tau$ are
\begin{equation}
    w_{\mathrm{I}k} \equiv \frac{1}{\sqrt{4\pi c |k|}} \exp[i(k\xi - |k|c\tau)].
\end{equation}
Due to the presence of the Rindler horizon, the region~II ($x < -c|t|$) Rindler modes are completely inaccessible to Rob and play no role in the analysis. 

\begin{figure}[ht]
    \centering
    \includegraphics[width=\linewidth]{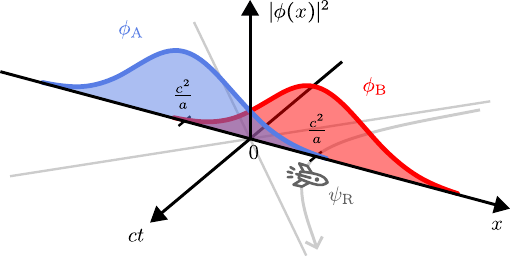}
    \caption{Two inertial observers, Alice and Bob, have access to orthogonal wave packets $\phi_\mathrm{A}$ and $\phi_\mathrm{B}$, respectively. At time $t = 0$, the envelopes of the wave packets are centered at positions $\pm c^2/a$ in Minkowski coordinates $(ct,x)$. Using these wave packets, either a two-mode squeezed state or two thermal states are prepared. An observer Rob, accelerating with constant proper acceleration $a$ in the $x$-direction, measures the state of the field by coupling to a wave packet $\psi_\mathrm{R}$ at $t = 0$ when his instantaneous velocity is zero.}
    \label{fig:2}
\end{figure}

Rob is in the possession of a detector that couples to a wave packet $\psi_\mathrm{R}(\xi,\tau)$ constructed solely from region~I positive-frequency Rindler modes. Under this assumption, the operator $\hat{d} = (\psi_\mathrm{R},\hat{\phi})$ satisfies the canonical commutation relations and annihilates the Rindler vacuum. Accordingly, $\hat{d}^{(\dag)}$ can be identified as the annihilation (creation) operator associated with Rob’s wave packet. Rob performs his measurement at time $t=0$, when his instantaneous velocity is zero. At this moment, we choose the envelopes of $\phi_\mathrm{A}$ and $\phi_\mathrm{B}$ to be centered symmetrically around positions $\pm c^2/a$ with respect to the event horizon (see Fig.~\ref{fig:2}). 

In the \emph{entangled} scenario, Alice and Bob, stationary with respect to each other, prepare a two-mode squeezed state:
\begin{equation}\label{TMSS}
    \hat{S}_\mathrm{AB} \ket{0}_\mathrm{M} = \exp[s(\hat{a}^\dag \hat{b}^\dag - \hat{a} \hat{b})] \ket{0}_\mathrm{M},
\end{equation}
where $s\geq0$ is the squeezing parameter, and the annihilation operators $\hat{a}$ and $\hat{b}$ are associated with the wave packets $\phi_\mathrm{A}$ and $\phi_\mathrm{B}$, respectively. The squeezing operator $\hat{S}_\mathrm{AB}$ can be disentangled \cite{Ferraro} to yield an explicit formula:
\begin{equation}
    \hat{S}_\mathrm{AB} \ket{0}_\mathrm{M} = \frac{1}{\cosh s} \sum_{n=0}^\infty \tanh^n s \ket{n}_\mathrm{A} \otimes \ket{n}_\mathrm{B}.
\end{equation}

If Alice and Bob cannot communicate and have access only to their respective wave packets, each can effectively describe the available part of the state~\eqref{TMSS} using the reduced density matrix:
\begin{equation}\label{TS}
    \rho =  \frac{1}{\cosh^2 s} \sum_{n=0}^\infty \tanh^{2n} s \ketbra{n},
\end{equation}
which corresponds to a thermal state. This motivates the definition of the \emph{separable} scenario, in which Alice and Bob prepare a product state:
\begin{equation}\label{product_TS}
    \rho_\mathrm{AB} = \rho_\mathrm{A} \otimes \rho_\mathrm{B},
\end{equation}
where $\rho_\mathrm{A}$ and $\rho_\mathrm{B}$ are thermal states of the form~\eqref{TS}. We stress that Eq.~\eqref{product_TS} is a density operator of the quantum field on the same fixed classical background as in the entangled case. Although it admits many ensemble decompositions, the probabilities of Rob's measurement outcomes depend only on the density matrix and not on a particular decomposition.

Since Rob is accelerating, relativistic effects in his subsystem must be taken into account. In particular, the Minkowski vacuum state undergoes a squeezing transformation \cite{Unruh1976}, which in general mixes field modes and introduces thermal noise. Moreover, Rob’s measurement capabilities are fundamentally limited by the presence of an event horizon, restricting his access to the field. As a result, the state accessible to Rob is a mixed Gaussian state obtained by tracing out all field modes orthogonal to the detected wave packet $\psi_\mathrm{R}$. An explicit evaluation of this trace is not required, since any Gaussian state is fully characterized by its covariance matrix~$\sigma$. It is therefore sufficient to measure two orthogonal quadratures of Rob’s wave packet, $\hat{x} = \frac{1}{\sqrt{2}} (\hat{d} + \hat{d}^\dag)$ and $\hat{p} = \frac{1}{\sqrt{2} i} (\hat{d} - \hat{d}^\dag)$, and evaluate the correlations between the measurement outcomes:
\begin{equation}\label{eq:correlations}
    \sigma_{ij} \equiv \langle \hat{X}_i \hat{X}_j + \hat{X}_j \hat{X}_i \rangle,
\end{equation}
where $\hat{\mathbf{X}} = (\hat{x}, \hat{p})$.

Details of the calculation of the covariance matrices for both scenarios are provided in the Appendix. In each case, the matrix elements depend on the following inner products:
\begin{align}
    &\alpha \equiv (\psi_\mathrm{R}, \phi_\mathrm{A}), \quad \alpha' \equiv (\psi_\mathrm{R}, \phi^*_\mathrm{A}), \\
    &\beta \equiv (\psi_\mathrm{R}, \phi_\mathrm{B}), \quad \beta' \equiv (\psi_\mathrm{R}, \phi_\mathrm{B}^*),
\end{align}
together with the average number of particles detected by a uniformly accelerated observer in the Minkowski vacuum:
\begin{equation}
    \langle \hat{n} \rangle_\mathrm{U} \equiv \int \der k\, \frac{|(\psi_\mathrm{R}, w_{\mathrm{I}k})|^2}{e^{2 \pi |k| c^2/a}-1}.
\end{equation}
The covariance matrix for the entangled scenario is
\begin{widetext}
    \begin{align}\label{eq:Rob}
        \sigma ={} &(1 + 2 \langle \hat{n} \rangle_\mathrm{U}) \mathds{1} + 2 \sinh^2 s 
            \begin{pmatrix}
                |\beta + \beta'^*|^2 + |\alpha + \alpha'^*|^2 & 2\Im(\beta \beta' + \alpha \alpha') \\
                2\Im(\beta \beta' + \alpha \alpha')  & |\beta - \beta'^*|^2 + |\alpha - \alpha'^*|^2
            \end{pmatrix}\nonumber\\
            &+ 2 \sinh 2s 
            \begin{pmatrix}
                -\Re[(\beta + \beta'^*)(\alpha + \alpha'^*)] & -\Im(\beta \alpha + \beta' \alpha')\\
                -\Im(\beta \alpha + \beta' \alpha') & 
                \Re[(\beta - \beta'^*)(\alpha - \alpha'^*)]
            \end{pmatrix}.
    \end{align}
    The covariance matrix for the separable scenario, denoted by $\sigma_\mathrm{s}$, retains the same local contributions but lacks the cross terms mixing the coefficients associated with Alice’s and Bob’s wave packets:
    \begin{align}\label{eq:Rob_C}
    \sigma_\mathrm{s} =
    \begin{aligned}[t]
        &(1 + 2 \langle \hat{n} \rangle_\mathrm{U}) \mathds{1} + 2 \sinh^2 s 
        \begin{pmatrix}
            |\beta + \beta'^*|^2 + |\alpha + \alpha'^*|^2 & 2\Im(\beta \beta' + \alpha \alpha') \\
            2\Im(\beta \beta' + \alpha \alpha')  & |\beta - \beta'^*|^2 + |\alpha - \alpha'^*|^2
        \end{pmatrix}.
    \end{aligned}
\end{align}
\end{widetext}
For the two preparations considered here, the only distinction between the covariance matrices therefore originates from the cross terms involving overlaps between Alice’s wave packet and the wave packet detected by Rob, encoded in the coefficients $\alpha$ and $\alpha'$. If Alice’s wave packet has no support in the region accessible to Rob, so that $\alpha = \alpha' = 0$, the two covariance matrices coincide and the scenarios become operationally indistinguishable. Any nonzero overlap, however small, introduces differences in the measured correlations and, in principle, enables Rob to detect the presence of entanglement. In this sense, the corresponding measurement scheme can be regarded as an entanglement witness for the particular family of states considered here.

\section{State discrimination}\label{sec:state_discrimination}

We have demonstrated that, for wave packets with nonvanishing asymptotic tails, the state accessible to Rob can differ between the separable and entangled scenarios specified above. Suppose now that Rob wishes to determine which of these two preparations was used. This task reduces to distinguishing between two equiprobable mixed quantum states. However, due to fundamental limitations imposed by quantum mechanics, nonorthogonal states cannot be perfectly distinguishable \cite{NielsenChuang2010}. To address this problem, various optimal measurement strategies have been developed, depending on the chosen criterion. The two most commonly used distinguishability techniques are: unambiguous state discrimination and minimum error state discrimination \cite{Bergou2004}. In what follows, we restrict ourselves to the latter approach.

To discriminate between two given states, we may apply an arbitrary quantum measurement to the system. Without loss of generality, we can restrict our attention to a dichotomic positive-operator-valued measure (POVM), whose outcome is a binary value distinguishing the two states. The minimum error probability $p_{\mathrm{e},\mathrm{min}}$ achievable by optimizing over all such measurements is given by the Helstrom bound, which depends on the trace distance between the corresponding density matrices \cite{Helstrom1976}. 

In general, obtaining an analytical expression for the trace distance is challenging, so the Helstrom bound is often approximated using other distinguishability measures. For our purposes, fidelity offers a convenient alternative \cite{Weedbrook2012}. For single-mode Gaussian states with identical first moments, described by covariance matrices $\sigma$ and $\sigma_\mathrm{s}$, the fidelity is given by \cite{Scutaru1998, Nha2005, Olivares2006, Holevo1975}:
\begin{equation}\label{eq:fidelity}
    F = \frac{2}{\sqrt{\Delta + \delta} - \sqrt{\delta}},
\end{equation}
where $\Delta = \det(\sigma + \sigma_\mathrm{s})$ and $\delta = (\det \sigma - 1)(\det \sigma_\mathrm{s} - 1)$. Based on the fidelity, we can define two bounds on the minimum error probability \cite{Fuchs_Graff1999}:
\begin{equation}
    F_- = \tfrac{1}{2} \left(1 - \sqrt{1 - F}\right), \quad F_+ = \tfrac{1}{2} \sqrt{F},
\end{equation}
which yield the following estimate:
\begin{equation}
    F_- \leq p_{\mathrm{e},\mathrm{min}} \leq F_+.
\end{equation}
When Rob’s proper acceleration vanishes and the wave packets $\phi_\mathrm{B}$ and $\psi_\mathrm{R}$ are chosen to match perfectly, we have $\beta = 1$, $\langle \hat{n} \rangle_\mathrm{U} = \beta' = \alpha = \alpha' = 0$ and then $F_- = F_+ = p_{\mathrm{e},\mathrm{min}} = \frac{1}{2}$. For nonzero proper accelerations, Eqs.~\eqref{eq:Rob}--\eqref{eq:Rob_C} can be substituted into Eq.~\eqref{eq:fidelity} and studied numerically. In order to perform this calculation, specific forms of $\phi_\mathrm{A}(x,t)$, $\phi_\mathrm{B}(x,t)$, and $\psi_\mathrm{R}(\xi,t)$ need to be chosen. Since we consider measurements carried out at $t = \tau = 0$, it is sufficient to specify the wave packets and their first derivatives at this time only.

To model Alice’s and Bob’s initial Minkowski wave packets, we adopt Gaussian profiles:
\begin{align}
    \phi_\pm(x,0) &= \frac{1}{\sqrt{N\sqrt{2\pi}}} \exp\!\left[-\frac{x^2}{L^2} \pm i \frac{N}{L}x \right],\\
    \partial_t \phi_\pm(x,0) &= - i \frac{Nc}{L} \phi_\pm(x,0),
\end{align}
where $L$ characterizes the spatial width of the wave packet, and $N/L$ is the central frequency. The sign determines the dominant propagation direction of the constituent plane waves: right-moving for $\phi_+$ and left-moving for $\phi_-$. To ensure orthogonality, we choose only left-moving waves for Alice and right-moving waves for Bob. In addition, a low-frequency cutoff is applied by removing all plane-wave components with $|k| < \Lambda$. This eliminates an infrared divergence associated with zero wave number but leaves the shape and localization of the wave packets near $t=0$ essentially unchanged.

Rob’s position at the time of measurement depends on his proper acceleration, $x(0) = \frac{c^2}{a}$. Accordingly, the wave packet associated with Bob must be appropriately translated and renormalized. From now on, we denote the normalization constants by $\mathcal{N}$ with the relevant subscript. In units where $c=1$, Bob's wave packet then takes the form
\begin{equation}
    \phi_\mathrm{B} = \mathcal{N}_\mathrm{B} \int_\Lambda^\infty \mathrm{d} l\, (u_l,\phi_+(x-\tfrac{1}{a}))\, u_l.
\end{equation}
We assume that Alice's wave packet is symmetrically centered with respect to the event horizon and is therefore given by
\begin{equation}
    \phi_\mathrm{A} = \mathcal{N}_\mathrm{A} \int_{-\infty}^{-\Lambda} \mathrm{d} l\, (u_l,\phi_-(x+\tfrac{1}{a}))\, u_l.
\end{equation}
Having specified the initial states, we turn to the choice of Rob’s detection wave packet. If it were to resemble Bob’s original Minkowski wave packet as closely as possible, it would take the form~\cite{Dragan2013}:
\begin{align}
    \psi_\mathrm{R} = \mathcal{N} \int_\Lambda^\infty \mathrm{d}k\, (w_{\mathrm{I}k}, \phi_\mathrm{B}) w_{\mathrm{I}k}.
\end{align}
However, in the accelerated frame, left- and right-moving Minkowski plane waves decompose exclusively into left- and right-moving Rindler plane waves, respectively~\cite{Crispino2008}. For this choice, one would thus obtain $\alpha = \alpha' = 0$, making the two scenarios operationally indistinguishable. 

To achieve nonvanishing overlaps with Alice’s wave packet, Rob’s detection mode must include both positive- and negative-momentum components. We therefore consider the wave packet:
\begin{align}
    \psi_\mathrm{R} = \mathcal{N}_\mathrm{R} &\Biggl( \int_{-\infty}^{-\Lambda} \mathrm{d}k\, (w_{\mathrm{I}k}, \phi_\mathrm{A}) w_{\mathrm{I}k} \nonumber\\
    &\quad + \int_\Lambda^\infty \mathrm{d}k\, (w_{\mathrm{I}k}, \phi_\mathrm{B}) w_{\mathrm{I}k} \Biggr),
\end{align}
which, while not necessarily optimal for the discrimination task, suffices to demonstrate a finite degree of distinguishability. With this choice, the relevant overlaps read
\begin{align}
    \beta &= \mathcal{N}_\mathrm{R} \int_\Lambda^\infty \mathrm{d}k\, |(w_{\mathrm{I}k}, \phi_\mathrm{B})|^2, \label{eq:beta}\\
    \beta' &= \mathcal{N}_\mathrm{R} \int_\Lambda^\infty \mathrm{d}k\, (w_{\mathrm{I}k}, \phi_\mathrm{B})^* (w_{\mathrm{I}k}, \phi_\mathrm{B}^*),\\
    \alpha &= \mathcal{N}_\mathrm{R} \int_{-\infty}^{-\Lambda} \mathrm{d}k\, |(w_{\mathrm{I}k}, \phi_\mathrm{A})|^2,\\
    \alpha' &= \mathcal{N}_\mathrm{R} \int_{-\infty}^{-\Lambda} \mathrm{d}k\, (w_{\mathrm{I}k}, \phi_\mathrm{A})^* (w_{\mathrm{I}k}, \phi_\mathrm{A}^*),
\end{align}
and the corresponding average number of particles detected by Rob in the Minkowski vacuum is
\begin{align}
    \langle \hat{n} \rangle_\mathrm{U} &= \mathcal{N}_\mathrm{R}^2 \Biggl( \int_{-\infty}^{-\Lambda} \mathrm{d}k\ \sinh^2 r_{|k|/a} \, |(w_{\mathrm{I}k}, \phi_\mathrm{A})|^2 \nonumber\\
    &\quad\quad\quad\quad + \int_\Lambda^\infty \mathrm{d}k\ \sinh^2 r_{|k|/a} \, |(w_{\mathrm{I}k}, \phi_\mathrm{B})|^2 \label{eq:n_U} \Biggr), 
\end{align}
where $r_\Omega = \mathrm{arctanh}(e^{-\pi\Omega})$.

Using the identity:
\begin{align}
    (u_l,\phi_\pm(x-x_0)) &= \frac{N + |l| L}{2\sqrt{|l|N\sqrt{2\pi}}} e^{-\frac{(lL \mp N)^2}{4} - ilx_0},
\end{align}
together with the transformation coefficients~\cite{Checinska2015}:
\begin{align}
    (w_{\mathrm{I}k}, u_l) &= \frac{i}{4 \pi} \frac{e^{\frac{\pi k}{2a}}}{\sqrt{|kl|}} \left(\frac{l}{a}\right)^{\frac{ik}{a}} \left(\frac{k}{|k|} + \frac{l}{|l|}\right) \Gamma\!\left(1-\frac{ik}{a}\right),\\
    (w_{\mathrm{I}k}, u_l^*) &= e^{-\frac{\pi |k|}{a}} (w_{\mathrm{I}k}, u_l),
\end{align}
and substituting them into the following expressions:
\begin{align}
    (w_{\mathrm{I}k}, \phi_\mathrm{B}) &= \mathcal{N}_\mathrm{B} \int_\Lambda^\infty \mathrm{d} l\, (u_l,\phi_+(x-\tfrac{1}{a})) (w_{\mathrm{I}k}, u_l),\\
    (w_{\mathrm{I}k}, \phi_\mathrm{B}^*) &= \mathcal{N}_\mathrm{B} \int_\Lambda^\infty \mathrm{d} l\, (u_l,\phi_+(x-\tfrac{1}{a}))^* (w_{\mathrm{I}k}, u_l^*),\\
    (w_{\mathrm{I}k}, \phi_\mathrm{A}) &= \mathcal{N}_\mathrm{A} \int_{-\infty}^{-\Lambda} \mathrm{d} l\, (u_l,\phi_-(x+\tfrac{1}{a})) (w_{\mathrm{I}k}, u_l),\\
    (w_{\mathrm{I}k}, \phi_\mathrm{A}^*) &= \mathcal{N}_\mathrm{A} \int_{-\infty}^{-\Lambda} \mathrm{d} l\, (u_l,\phi_-(x+\tfrac{1}{a}))^* (w_{\mathrm{I}k}, u_l^*),
\end{align}
one can numerically compute the quantities appearing in Eqs.~\eqref{eq:beta}--\eqref{eq:n_U}, which are necessary to determine the covariance matrices.

\begin{figure}[]
    \centering
    \includegraphics[width=\linewidth]{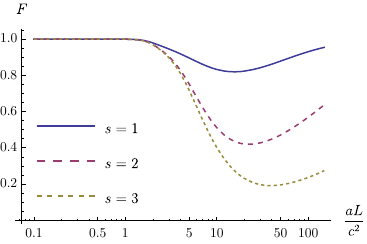}
    \caption{Fidelity $F$ as a function of the dimensionless acceleration parameter $\frac{aL}{c^2}$ for $N = 6$, $\Lambda = \frac{c}{2L}$ and different values of $s$.}
    \label{fig:fidelity_plot_s}
\end{figure}
\begin{figure}[]
    \centering
    \includegraphics[width=\linewidth]{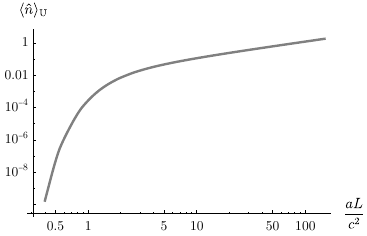}
    \caption{Average number of particles seen by an accelerated detector in the vacuum $\langle \hat{n} \rangle_\mathrm{U}$ as a function of the dimensionless acceleration parameter $\frac{aL}{c^2}$ for $N = 6$ and $\Lambda = \frac{c}{2L}$.}
    \label{fig:particles_plot}
\end{figure}

In Fig.~\ref{fig:fidelity_plot_s}, we present the numerically calculated fidelity between the covariance matrices corresponding to the separable and entangled scenarios for different values of the squeezing parameter~$s$. In the low-acceleration regime, $\frac{aL}{c^2} \ll 1$, the portions of the initial wave packets extending beyond the event horizon are negligible. As a result, the overlap between Alice’s and Rob’s wave packets effectively vanishes, leading to minimal distinguishability. This situation closely approximates that in which the wave packets associated with Alice and Bob are localized and spatially separated.

As the acceleration increases, however, the contribution of the wave-packet tails that cross the horizon becomes significant. In this regime, the fidelity decreases, reaches a well-defined minimum, and eventually returns to unity for sufficiently large accelerations. Moreover, the curves corresponding to different squeezing parameters, which initially overlap, begin to separate: larger initial squeezing shifts the minimum to higher accelerations and leads to a lower minimal fidelity.

This behavior can be attributed to the effect of Unruh particles detected by Rob~\eqref{eq:n_U}, which appear even when the state contains no excitations ($s = 0$) and correspond, in the Schwarzschild picture, to Hawking radiation. The expected number of such particles per Rindler frequency follows a Bose--Einstein distribution at the Unruh temperature:
\begin{equation}
    \langle \hat{n}_k \rangle = \frac{1}{e^{2 \pi |k| c^2/a}-1}.
\end{equation}
These thermal excitations are concentrated predominantly at frequencies below a critical threshold $k_c = \frac{a}{2 \pi c^2}$. For sufficiently high accelerations (or, equivalently, when the wave packets lie close to the event horizon) this critical frequency exceeds the imposed low-frequency cutoff~$\Lambda$. In this regime, the contribution of low Rindler frequencies dominates the mode decomposition. Consequently, the average number of Unruh particles, $\langle \hat{n} \rangle_\mathrm{U}$, increases and asymptotically follows a power-law dependence, as illustrated in Fig.~\ref{fig:particles_plot}.

At sufficiently high accelerations, acceleration-induced thermal noise amplifies the components of the covariance matrices proportional to the identity, effectively suppressing the relative contribution of the remaining entries. Consequently, the covariance matrices for the separable and entangled scenarios gradually become indistinguishable. This loss of distinguishability can be delayed to higher accelerations by increasing the squeezing parameter $s$, which counteracts the growing $\langle \hat{n} \rangle_\mathrm{U}$.

Using the fidelity, one can compute bounds on the minimum error probability $p_\mathrm{e,min}$ in the state discrimination task. As an illustration, Fig.~\ref{fig:p_e,min_logplot} shows both bounds for a squeezing parameter $s = 3$. In the optimal configuration, the minimum error probability satisfies $p_\mathrm{e,min} \lesssim 0.22$. This bound can be further improved by increasing $s$. In the large-acceleration limit, the minimum error probability is expected to asymptotically approach $p_\mathrm{e,min} = \frac{1}{2}$.

\begin{figure}[]
    \centering
    \includegraphics[width=\linewidth]{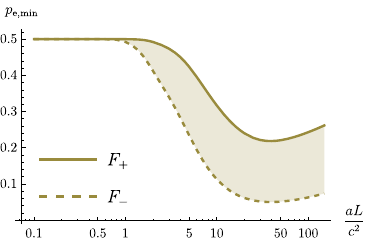}
    \caption{Bounds on the minimum error probability, $F_- \leq p_\mathrm{e,min} \leq F_+$, as a function of the dimensionless acceleration parameter $\frac{aL}{c^2}$ for $N = 6$, $\Lambda = \frac{c}{2L}$ and $s = 3$.}
    \label{fig:p_e,min_logplot}
\end{figure}

We have thus demonstrated that, contrary to common intuition, a uniformly accelerated observer, Rob, can in principle discriminate between the entangled and separable preparations considered in this work, even when the corresponding wave packets are centered on opposite sides of an event horizon. The minimum error probability in the corresponding discrimination task depends on the spatial separation of the wave-packet centers from the horizon and the squeezing parameter~$s$.

\section{Conclusions}\label{sec:conclusions}

Employing the equivalence principle within the framework of quantum field theory in curved spacetime, we addressed the problem of distinguishing between separable and entangled states of two wave packets centered on opposite sides of a black hole event horizon. In the near-horizon regime, the external stationary observer was modeled as a uniformly accelerated observer, Rob, while the corresponding wave packets were taken to be prepared by inertial observers, Alice and Bob. We found that, assuming the initial wave packets are constructed solely from positive-frequency Minkowski plane waves, the two scenarios can be, in principle, distinguishable. This distinguishability arises from fundamental limitations on the localizability of quantum wave packets. As demonstrated in a concrete realization, the effect becomes appreciable only when the asymptotic tails of the wave packets extend across the event horizon and vanishes in the infinite acceleration limit. For wave packets localized farther from the horizon the distinguishability is progressively suppressed, with the minimum error probability in the corresponding discrimination task remaining infinitesimally below its maximal value.

Our results suggest that fundamental theorems governing the localization properties of quantum states may allow for a limited form of information accessibility across an event horizon. While this does not imply any violation of causality or signal transmission, it demonstrates that quantum correlations can leave observable imprints even when one of the correlated wave packets lies predominantly beyond the horizon. Taken together with related work~\cite{Falcone2023}, these observations indicate that certain aspects of the black hole information problem may be reconsidered from an alternative perspective.

\begin{acknowledgments}
We thank Sir Roger Penrose for an interesting discussion about this problem.
\end{acknowledgments}

\section*{Data availability}
The data that support the findings of this article are openly available~\cite{Michalski2026}.

\appendix

\section{Calculation of the~covariance matrices}
In this appendix, we present the detailed derivation of the covariance matrices \eqref{eq:Rob} and \eqref{eq:Rob_C}. In both cases, the matrices are $2 \times 2$ dimensional and their elements are obtained by evaluating Eq.~\eqref{eq:correlations}. To facilitate the calculation, it is convenient to decompose the operator $\hat{d}$ as
\begin{align}\label{eq:decomposition}
    \hat{d} ={} & \alpha \hat{a} + \alpha' \hat{a}^\dag + \beta \hat{b} + \beta' \hat{b}^\dag \nonumber\\ 
    & + (\hat{d} - \alpha \hat{a} - \alpha' \hat{a}^\dag - \beta \hat{b} - \beta' \hat{b}^\dag).
\end{align}
The expression in parentheses contains no contribution from the operators $\hat{a}$, $\hat{a}^\dag$, $\hat{b}$, or $\hat{b}^\dag$. For brevity, we define
\begin{equation}
    \hat{d}_\perp \equiv \hat{d} - \alpha \hat{a} - \alpha' \hat{a}^\dag - \beta \hat{b} - \beta' \hat{b}^\dag.
\end{equation}
The covariance matrix for the entangled scenario in Eq.~\eqref{eq:Rob} follows from this decomposition by a direct extension of the derivation given in the Appendix of Ref.~\cite{Dragan2013}.

We now focus on the separable scenario. The field is initially prepared as a product of thermal states~\eqref{product_TS}. The wave packets associated with the operators $\hat{a}$, $\hat{b}$, and $\hat{d}_\perp$ are mutually orthogonal. Since a thermal state is Gaussian and diagonal in the Fock basis, all its first moments vanish. The only nonzero second moments of the initial states prepared by Alice and Bob are
\begin{align}\label{eq:TS_moments}
    &\Tr{\hat{a}^\dag \hat{a}\, \rho_\mathrm{A}} = \Tr\big\{\hat{b}^\dag \hat{b}\, \rho_\mathrm{B}\big\} = \sinh^2 s,\\
    &\Tr{\hat{a} \hat{a}^\dag \rho_\mathrm{A}} = \Tr\big\{\hat{b} \hat{b}^\dag \rho_\mathrm{B}\big\} = 1 + \sinh^2 s.
\end{align}
Using these relations, each element of the covariance matrix can be evaluated explicitly. As an illustration, consider the $\sigma_{11}$ component:
\begin{equation}
    \sigma_{11} = \Tr{(\hat{d} + \hat{d}^\dag)^2 \rho_\mathrm{AB}}.
\end{equation}
Applying Eq.~\eqref{eq:decomposition} and discarding vanishing expectation values yields
\begin{align}
    \sigma_{11} ={} & \left(|\beta + \beta'^*|^2 + |\alpha + \alpha'^*|^2 \right) \left(1 + 2 \sinh^2 s \right)\nonumber\\
    &+ \Tr{ (\hat{d}_\perp + \hat{d}_\perp^\dag)^2 \rho_\mathrm{AB}}.
\end{align}
From the orthogonality of the wave packets it follows that the last term is identical to the expectation value in the Minkowski vacuum:
\begin{equation}
     \Tr{ (\hat{d}_\perp + \hat{d}_\perp^\dag)^2 \rho_\mathrm{AB}} = \prescript{}{\mathrm{M}}{\bra{0}}  (\hat{d}_\perp + \hat{d}_\perp^\dag)^2 \ket{0}_\mathrm{M}.
\end{equation}
Expanding $\hat{d}_\perp$ and noting that $\hat{a}$ and $\hat{b}$ annihilate the Minkowski vacuum, we obtain
\begin{align}\label{eq:Rob_C_intermediate}
    \sigma_{11} ={} & \left(|\beta + \beta'^*|^2 + |\alpha + \alpha'^*|^2 \right) \left(2 + 2 \sinh^2 s \right)\nonumber\\
    &- 2\Re\!\left[(\beta + \beta'^*) \prescript{}{\mathrm{M}}{\bra{0}} \hat{b} (\hat{d} + \hat{d}^\dag) \ket{0}_\mathrm{M}\right] \nonumber\\
    &- 2\Re\!\left[(\alpha + \alpha'^*) \prescript{}{\mathrm{M}}{\bra{0}} \hat{a} (\hat{d} + \hat{d}^\dag) \ket{0}_\mathrm{M}\right] \nonumber\\
    & + \prescript{}{\mathrm{M}}{\bra{0}} (\hat{d} + \hat{d}^\dag)^2 \ket{0}_\mathrm{M}.
\end{align}
The last term can be calculated by expressing the Minkowski vacuum in terms of the Rindler vacuum~\cite{Dragan2013}, giving
\begin{equation}\label{eq:vac_exp_d}
    \prescript{}{\mathrm{M}}{\bra{0}} (\hat{d} + \hat{d}^\dag)^2 \ket{0}_\mathrm{M} = 1 + 2 \langle \hat{n} \rangle_\mathrm{U}.
\end{equation}
Furthermore, evaluating the mixed terms through commutators leads to
\begin{align}
    &\prescript{}{\mathrm{M}}{\bra{0}} \hat{b} (\hat{d} + \hat{d}^\dag) \ket{0}_\mathrm{M} = \prescript{}{\mathrm{M}}{\bra{0}} [\hat{b}, \hat{d} + \hat{d}^\dag] \ket{0}_\mathrm{M} = \beta^* + \beta', \label{eq:vac_exp_mix_B}\\
    &\prescript{}{\mathrm{M}}{\bra{0}} \hat{a} (\hat{d} + \hat{d}^\dag) \ket{0}_\mathrm{M} = \prescript{}{\mathrm{M}}{\bra{0}} [\hat{a}, \hat{d} + \hat{d}^\dag] \ket{0}_\mathrm{M} = \alpha^* + \alpha'. \label{eq:vac_exp_mix_A}
\end{align}
Finally, substituting Eqs.~\eqref{eq:vac_exp_d}--\eqref{eq:vac_exp_mix_A} into Eq.~\eqref{eq:Rob_C_intermediate} and simplifying, we find
\begin{align}
    \sigma_{11} ={} & 2 \left(|\beta + \beta'^*|^2 + |\alpha + \alpha'^*|^2 \right) \sinh^2 s \nonumber\\
    &+ 1 + 2 \langle \hat{n} \rangle_\mathrm{U}.
\end{align}
The remaining matrix elements can be calculated in the same manner, resulting in Eq.~\eqref{eq:Rob_C}.

\bibliography{bibliography}

\begin{thebibliography}{43}%
\makeatletter
\providecommand \@ifxundefined [1]{%
 \@ifx{#1\undefined}
}%
\providecommand \@ifnum [1]{%
 \ifnum #1\expandafter \@firstoftwo
 \else \expandafter \@secondoftwo
 \fi
}%
\providecommand \@ifx [1]{%
 \ifx #1\expandafter \@firstoftwo
 \else \expandafter \@secondoftwo
 \fi
}%
\providecommand \natexlab [1]{#1}%
\providecommand \enquote  [1]{``#1''}%
\providecommand \bibnamefont  [1]{#1}%
\providecommand \bibfnamefont [1]{#1}%
\providecommand \citenamefont [1]{#1}%
\providecommand \href@noop [0]{\@secondoftwo}%
\providecommand \href [0]{\begingroup \@sanitize@url \@href}%
\providecommand \@href[1]{\@@startlink{#1}\@@href}%
\providecommand \@@href[1]{\endgroup#1\@@endlink}%
\providecommand \@sanitize@url [0]{\catcode `\\12\catcode `\$12\catcode `\&12\catcode `\#12\catcode `\^12\catcode `\_12\catcode `\%12\relax}%
\providecommand \@@startlink[1]{}%
\providecommand \@@endlink[0]{}%
\providecommand \url  [0]{\begingroup\@sanitize@url \@url }%
\providecommand \@url [1]{\endgroup\@href {#1}{\urlprefix }}%
\providecommand \urlprefix  [0]{URL }%
\providecommand \Eprint [0]{\href }%
\providecommand \doibase [0]{https://doi.org/}%
\providecommand \selectlanguage [0]{\@gobble}%
\providecommand \bibinfo  [0]{\@secondoftwo}%
\providecommand \bibfield  [0]{\@secondoftwo}%
\providecommand \translation [1]{[#1]}%
\providecommand \BibitemOpen [0]{}%
\providecommand \bibitemStop [0]{}%
\providecommand \bibitemNoStop [0]{.\EOS\space}%
\providecommand \EOS [0]{\spacefactor3000\relax}%
\providecommand \BibitemShut  [1]{\csname bibitem#1\endcsname}%
\let\auto@bib@innerbib\@empty
\bibitem [{\citenamefont {Wald}(1984)}]{Wald1984}%
  \BibitemOpen
  \bibfield  {author} {\bibinfo {author} {\bibfnamefont {R.~M.}\ \bibnamefont {Wald}},\ }\href {https://doi.org/10.7208/chicago/9780226870373.001.0001} {\emph {\bibinfo {title} {{General Relativity}}}}\ (\bibinfo  {publisher} {Chicago University Press},\ \bibinfo {address} {Chicago},\ \bibinfo {year} {1984})\BibitemShut {NoStop}%
\bibitem [{\citenamefont {Curiel}(2019)}]{Curiel2019}%
  \BibitemOpen
  \bibfield  {author} {\bibinfo {author} {\bibfnamefont {E.}~\bibnamefont {Curiel}},\ }\bibfield  {title} {\bibinfo {title} {The many definitions of a black hole},\ }\href {https://doi.org/10.1038/s41550-018-0602-1} {\bibfield  {journal} {\bibinfo  {journal} {Nat. Astron.}\ }\textbf {\bibinfo {volume} {3}},\ \bibinfo {pages} {27} (\bibinfo {year} {2019})}\BibitemShut {NoStop}%
\bibitem [{\citenamefont {Hawking}(1974)}]{Hawking1974}%
  \BibitemOpen
  \bibfield  {author} {\bibinfo {author} {\bibfnamefont {S.~W.}\ \bibnamefont {Hawking}},\ }\bibfield  {title} {\bibinfo {title} {Black hole explosions?},\ }\href {https://doi.org/10.1038/248030a0} {\bibfield  {journal} {\bibinfo  {journal} {Nature}\ }\textbf {\bibinfo {volume} {248}},\ \bibinfo {pages} {30} (\bibinfo {year} {1974})}\BibitemShut {NoStop}%
\bibitem [{\citenamefont {Hawking}(1975)}]{Hawking1975}%
  \BibitemOpen
  \bibfield  {author} {\bibinfo {author} {\bibfnamefont {S.~W.}\ \bibnamefont {Hawking}},\ }\bibfield  {title} {\bibinfo {title} {Particle creation by black holes},\ }\href {https://doi.org/10.1007/BF02345020} {\bibfield  {journal} {\bibinfo  {journal} {Commun. Math. Phys.}\ }\textbf {\bibinfo {volume} {43}},\ \bibinfo {pages} {199} (\bibinfo {year} {1975})}\BibitemShut {NoStop}%
\bibitem [{\citenamefont {Wald}(1994)}]{Wald1994}%
  \BibitemOpen
  \bibfield  {author} {\bibinfo {author} {\bibfnamefont {R.~M.}\ \bibnamefont {Wald}},\ }\href@noop {} {\emph {\bibinfo {title} {Quantum Field Theory in Curved Spacetime and Black Hole Thermodynamics}}}\ (\bibinfo  {publisher} {The University of Chicago Press},\ \bibinfo {year} {1994})\BibitemShut {NoStop}%
\bibitem [{\citenamefont {Birrell}\ and\ \citenamefont {Davies}(1982)}]{Birrell1982}%
  \BibitemOpen
  \bibfield  {author} {\bibinfo {author} {\bibfnamefont {N.~D.}\ \bibnamefont {Birrell}}\ and\ \bibinfo {author} {\bibfnamefont {P.~C.~W.}\ \bibnamefont {Davies}},\ }\href {https://doi.org/10.1017/CBO9780511622632} {\emph {\bibinfo {title} {{Quantum Fields in Curved Space}}}},\ Cambridge Monographs on Mathematical Physics\ (\bibinfo  {publisher} {Cambridge University Press},\ \bibinfo {address} {Cambridge},\ \bibinfo {year} {1982})\BibitemShut {NoStop}%
\bibitem [{\citenamefont {Alsing}\ and\ \citenamefont {Milburn}(2003)}]{Alsing2003}%
  \BibitemOpen
  \bibfield  {author} {\bibinfo {author} {\bibfnamefont {P.~M.}\ \bibnamefont {Alsing}}\ and\ \bibinfo {author} {\bibfnamefont {G.~J.}\ \bibnamefont {Milburn}},\ }\bibfield  {title} {\bibinfo {title} {Teleportation with a uniformly accelerated partner},\ }\href {https://doi.org/10.1103/PhysRevLett.91.180404} {\bibfield  {journal} {\bibinfo  {journal} {Phys. Rev. Lett.}\ }\textbf {\bibinfo {volume} {91}},\ \bibinfo {pages} {180404} (\bibinfo {year} {2003})}\BibitemShut {NoStop}%
\bibitem [{\citenamefont {Fuentes-Schuller}\ and\ \citenamefont {Mann}(2005)}]{Fuentes2005}%
  \BibitemOpen
  \bibfield  {author} {\bibinfo {author} {\bibfnamefont {I.}~\bibnamefont {Fuentes-Schuller}}\ and\ \bibinfo {author} {\bibfnamefont {R.~B.}\ \bibnamefont {Mann}},\ }\bibfield  {title} {\bibinfo {title} {Alice falls into a black hole: Entanglement in noninertial frames},\ }\href {https://doi.org/10.1103/PhysRevLett.95.120404} {\bibfield  {journal} {\bibinfo  {journal} {Phys. Rev. Lett.}\ }\textbf {\bibinfo {volume} {95}},\ \bibinfo {pages} {120404} (\bibinfo {year} {2005})}\BibitemShut {NoStop}%
\bibitem [{\citenamefont {Alsing}\ \emph {et~al.}(2006)\citenamefont {Alsing}, \citenamefont {Fuentes-Schuller}, \citenamefont {Mann},\ and\ \citenamefont {Tessier}}]{Alsing2006}%
  \BibitemOpen
  \bibfield  {author} {\bibinfo {author} {\bibfnamefont {P.~M.}\ \bibnamefont {Alsing}}, \bibinfo {author} {\bibfnamefont {I.}~\bibnamefont {Fuentes-Schuller}}, \bibinfo {author} {\bibfnamefont {R.~B.}\ \bibnamefont {Mann}},\ and\ \bibinfo {author} {\bibfnamefont {T.~E.}\ \bibnamefont {Tessier}},\ }\bibfield  {title} {\bibinfo {title} {Entanglement of {D}irac fields in noninertial frames},\ }\href {https://doi.org/10.1103/PhysRevA.74.032326} {\bibfield  {journal} {\bibinfo  {journal} {Phys. Rev. A}\ }\textbf {\bibinfo {volume} {74}},\ \bibinfo {pages} {032326} (\bibinfo {year} {2006})}\BibitemShut {NoStop}%
\bibitem [{\citenamefont {Mart\'{\i}n-Mart\'{\i}nez}\ and\ \citenamefont {Le\'on}(2009)}]{Martin-Martinez2009}%
  \BibitemOpen
  \bibfield  {author} {\bibinfo {author} {\bibfnamefont {E.}~\bibnamefont {Mart\'{\i}n-Mart\'{\i}nez}}\ and\ \bibinfo {author} {\bibfnamefont {J.}~\bibnamefont {Le\'on}},\ }\bibfield  {title} {\bibinfo {title} {Fermionic entanglement that survives a black hole},\ }\href {https://doi.org/10.1103/PhysRevA.80.042318} {\bibfield  {journal} {\bibinfo  {journal} {Phys. Rev. A}\ }\textbf {\bibinfo {volume} {80}},\ \bibinfo {pages} {042318} (\bibinfo {year} {2009})}\BibitemShut {NoStop}%
\bibitem [{\citenamefont {Mart\'{\i}n-Mart\'{\i}nez}\ \emph {et~al.}(2010)\citenamefont {Mart\'{\i}n-Mart\'{\i}nez}, \citenamefont {Garay},\ and\ \citenamefont {Le\'on}}]{Martin-Martinez2010}%
  \BibitemOpen
  \bibfield  {author} {\bibinfo {author} {\bibfnamefont {E.}~\bibnamefont {Mart\'{\i}n-Mart\'{\i}nez}}, \bibinfo {author} {\bibfnamefont {L.~J.}\ \bibnamefont {Garay}},\ and\ \bibinfo {author} {\bibfnamefont {J.}~\bibnamefont {Le\'on}},\ }\bibfield  {title} {\bibinfo {title} {Unveiling quantum entanglement degradation near a {S}chwarzschild black hole},\ }\href {https://doi.org/10.1103/PhysRevD.82.064006} {\bibfield  {journal} {\bibinfo  {journal} {Phys. Rev. D}\ }\textbf {\bibinfo {volume} {82}},\ \bibinfo {pages} {064006} (\bibinfo {year} {2010})}\BibitemShut {NoStop}%
\bibitem [{\citenamefont {Mart\'{\i}n-Mart\'{\i}nez}\ and\ \citenamefont {Le\'on}(2010)}]{Martin-Martinez_Leon2010}%
  \BibitemOpen
  \bibfield  {author} {\bibinfo {author} {\bibfnamefont {E.}~\bibnamefont {Mart\'{\i}n-Mart\'{\i}nez}}\ and\ \bibinfo {author} {\bibfnamefont {J.}~\bibnamefont {Le\'on}},\ }\bibfield  {title} {\bibinfo {title} {Quantum correlations through event horizons: {F}ermionic versus bosonic entanglement},\ }\href {https://doi.org/10.1103/PhysRevA.81.032320} {\bibfield  {journal} {\bibinfo  {journal} {Phys. Rev. A}\ }\textbf {\bibinfo {volume} {81}},\ \bibinfo {pages} {032320} (\bibinfo {year} {2010})}\BibitemShut {NoStop}%
\bibitem [{\citenamefont {Bruschi}\ \emph {et~al.}(2010)\citenamefont {Bruschi}, \citenamefont {Louko}, \citenamefont {Mart\'{\i}n-Mart\'{\i}nez}, \citenamefont {Dragan},\ and\ \citenamefont {Fuentes}}]{Bruschi2010}%
  \BibitemOpen
  \bibfield  {author} {\bibinfo {author} {\bibfnamefont {D.~E.}\ \bibnamefont {Bruschi}}, \bibinfo {author} {\bibfnamefont {J.}~\bibnamefont {Louko}}, \bibinfo {author} {\bibfnamefont {E.}~\bibnamefont {Mart\'{\i}n-Mart\'{\i}nez}}, \bibinfo {author} {\bibfnamefont {A.}~\bibnamefont {Dragan}},\ and\ \bibinfo {author} {\bibfnamefont {I.}~\bibnamefont {Fuentes}},\ }\bibfield  {title} {\bibinfo {title} {Unruh effect in quantum information beyond the single-mode approximation},\ }\href {https://doi.org/10.1103/PhysRevA.82.042332} {\bibfield  {journal} {\bibinfo  {journal} {Phys. Rev. A}\ }\textbf {\bibinfo {volume} {82}},\ \bibinfo {pages} {042332} (\bibinfo {year} {2010})}\BibitemShut {NoStop}%
\bibitem [{\citenamefont {Dragan}\ \emph {et~al.}(2013{\natexlab{a}})\citenamefont {Dragan}, \citenamefont {Doukas}, \citenamefont {Martín-Martínez},\ and\ \citenamefont {Bruschi}}]{Dragan_Doukas2013}%
  \BibitemOpen
  \bibfield  {author} {\bibinfo {author} {\bibfnamefont {A.}~\bibnamefont {Dragan}}, \bibinfo {author} {\bibfnamefont {J.}~\bibnamefont {Doukas}}, \bibinfo {author} {\bibfnamefont {E.}~\bibnamefont {Martín-Martínez}},\ and\ \bibinfo {author} {\bibfnamefont {D.~E.}\ \bibnamefont {Bruschi}},\ }\bibfield  {title} {\bibinfo {title} {Localized projective measurement of a quantum field in non-inertial frames},\ }\href {https://doi.org/10.1088/0264-9381/30/23/235006} {\bibfield  {journal} {\bibinfo  {journal} {Class. Quantum Grav.}\ }\textbf {\bibinfo {volume} {30}},\ \bibinfo {pages} {235006} (\bibinfo {year} {2013}{\natexlab{a}})}\BibitemShut {NoStop}%
\bibitem [{\citenamefont {Downes}\ \emph {et~al.}(2011)\citenamefont {Downes}, \citenamefont {Fuentes},\ and\ \citenamefont {Ralph}}]{Downes2011}%
  \BibitemOpen
  \bibfield  {author} {\bibinfo {author} {\bibfnamefont {T.~G.}\ \bibnamefont {Downes}}, \bibinfo {author} {\bibfnamefont {I.}~\bibnamefont {Fuentes}},\ and\ \bibinfo {author} {\bibfnamefont {T.~C.}\ \bibnamefont {Ralph}},\ }\bibfield  {title} {\bibinfo {title} {Entangling moving cavities in noninertial frames},\ }\href {https://doi.org/10.1103/PhysRevLett.106.210502} {\bibfield  {journal} {\bibinfo  {journal} {Phys. Rev. Lett.}\ }\textbf {\bibinfo {volume} {106}},\ \bibinfo {pages} {210502} (\bibinfo {year} {2011})}\BibitemShut {NoStop}%
\bibitem [{\citenamefont {Bruschi}\ \emph {et~al.}(2012)\citenamefont {Bruschi}, \citenamefont {Fuentes},\ and\ \citenamefont {Louko}}]{Bruschi2012}%
  \BibitemOpen
  \bibfield  {author} {\bibinfo {author} {\bibfnamefont {D.~E.}\ \bibnamefont {Bruschi}}, \bibinfo {author} {\bibfnamefont {I.}~\bibnamefont {Fuentes}},\ and\ \bibinfo {author} {\bibfnamefont {J.}~\bibnamefont {Louko}},\ }\bibfield  {title} {\bibinfo {title} {Voyage to {A}lpha {C}entauri: {E}ntanglement degradation of cavity modes due to motion},\ }\href {https://doi.org/10.1103/PhysRevD.85.061701} {\bibfield  {journal} {\bibinfo  {journal} {Phys. Rev. D}\ }\textbf {\bibinfo {volume} {85}},\ \bibinfo {pages} {061701} (\bibinfo {year} {2012})}\BibitemShut {NoStop}%
\bibitem [{\citenamefont {Friis}\ \emph {et~al.}(2012)\citenamefont {Friis}, \citenamefont {Bruschi}, \citenamefont {Louko},\ and\ \citenamefont {Fuentes}}]{Friis2012}%
  \BibitemOpen
  \bibfield  {author} {\bibinfo {author} {\bibfnamefont {N.}~\bibnamefont {Friis}}, \bibinfo {author} {\bibfnamefont {D.~E.}\ \bibnamefont {Bruschi}}, \bibinfo {author} {\bibfnamefont {J.}~\bibnamefont {Louko}},\ and\ \bibinfo {author} {\bibfnamefont {I.}~\bibnamefont {Fuentes}},\ }\bibfield  {title} {\bibinfo {title} {Motion generates entanglement},\ }\href {https://doi.org/10.1103/PhysRevD.85.081701} {\bibfield  {journal} {\bibinfo  {journal} {Phys. Rev. D}\ }\textbf {\bibinfo {volume} {85}},\ \bibinfo {pages} {081701} (\bibinfo {year} {2012})}\BibitemShut {NoStop}%
\bibitem [{\citenamefont {Lin}\ and\ \citenamefont {Hu}(2008)}]{Lin2008}%
  \BibitemOpen
  \bibfield  {author} {\bibinfo {author} {\bibfnamefont {S.-Y.}\ \bibnamefont {Lin}}\ and\ \bibinfo {author} {\bibfnamefont {B.~L.}\ \bibnamefont {Hu}},\ }\bibfield  {title} {\bibinfo {title} {Entanglement, recoherence and information flow in an accelerated detector—quantum field system: implications for the black hole information issue},\ }\href {https://doi.org/10.1088/0264-9381/25/15/154004} {\bibfield  {journal} {\bibinfo  {journal} {Class. Quantum Grav.}\ }\textbf {\bibinfo {volume} {25}},\ \bibinfo {pages} {154004} (\bibinfo {year} {2008})}\BibitemShut {NoStop}%
\bibitem [{\citenamefont {Lin}\ \emph {et~al.}(2008)\citenamefont {Lin}, \citenamefont {Chou},\ and\ \citenamefont {Hu}}]{Lin_Chung2008}%
  \BibitemOpen
  \bibfield  {author} {\bibinfo {author} {\bibfnamefont {S.-Y.}\ \bibnamefont {Lin}}, \bibinfo {author} {\bibfnamefont {C.-H.}\ \bibnamefont {Chou}},\ and\ \bibinfo {author} {\bibfnamefont {B.~L.}\ \bibnamefont {Hu}},\ }\bibfield  {title} {\bibinfo {title} {Disentanglement of two harmonic oscillators in relativistic motion},\ }\href {https://doi.org/10.1103/PhysRevD.78.125025} {\bibfield  {journal} {\bibinfo  {journal} {Phys. Rev. D}\ }\textbf {\bibinfo {volume} {78}},\ \bibinfo {pages} {125025} (\bibinfo {year} {2008})}\BibitemShut {NoStop}%
\bibitem [{\citenamefont {Lin}\ and\ \citenamefont {Hu}(2009)}]{Lin2009}%
  \BibitemOpen
  \bibfield  {author} {\bibinfo {author} {\bibfnamefont {S.-Y.}\ \bibnamefont {Lin}}\ and\ \bibinfo {author} {\bibfnamefont {B.~L.}\ \bibnamefont {Hu}},\ }\bibfield  {title} {\bibinfo {title} {Temporal and spatial dependence of quantum entanglement from a field theory perspective},\ }\href {https://doi.org/10.1103/PhysRevD.79.085020} {\bibfield  {journal} {\bibinfo  {journal} {Phys. Rev. D}\ }\textbf {\bibinfo {volume} {79}},\ \bibinfo {pages} {085020} (\bibinfo {year} {2009})}\BibitemShut {NoStop}%
\bibitem [{\citenamefont {Doukas}\ and\ \citenamefont {Carson}(2010)}]{Doukas2010}%
  \BibitemOpen
  \bibfield  {author} {\bibinfo {author} {\bibfnamefont {J.}~\bibnamefont {Doukas}}\ and\ \bibinfo {author} {\bibfnamefont {B.}~\bibnamefont {Carson}},\ }\bibfield  {title} {\bibinfo {title} {Entanglement of two qubits in a relativistic orbit},\ }\href {https://doi.org/10.1103/PhysRevA.81.062320} {\bibfield  {journal} {\bibinfo  {journal} {Phys. Rev. A}\ }\textbf {\bibinfo {volume} {81}},\ \bibinfo {pages} {062320} (\bibinfo {year} {2010})}\BibitemShut {NoStop}%
\bibitem [{\citenamefont {Dragan}\ \emph {et~al.}(2013{\natexlab{b}})\citenamefont {Dragan}, \citenamefont {Doukas},\ and\ \citenamefont {Mart\'{\i}n-Mart\'{\i}nez}}]{Dragan2013}%
  \BibitemOpen
  \bibfield  {author} {\bibinfo {author} {\bibfnamefont {A.}~\bibnamefont {Dragan}}, \bibinfo {author} {\bibfnamefont {J.}~\bibnamefont {Doukas}},\ and\ \bibinfo {author} {\bibfnamefont {E.}~\bibnamefont {Mart\'{\i}n-Mart\'{\i}nez}},\ }\bibfield  {title} {\bibinfo {title} {Localized detection of quantum entanglement through the event horizon},\ }\href {https://doi.org/10.1103/PhysRevA.87.052326} {\bibfield  {journal} {\bibinfo  {journal} {Phys. Rev. A}\ }\textbf {\bibinfo {volume} {87}},\ \bibinfo {pages} {052326} (\bibinfo {year} {2013}{\natexlab{b}})}\BibitemShut {NoStop}%
\bibitem [{\citenamefont {Richter}\ \emph {et~al.}(2017)\citenamefont {Richter}, \citenamefont {Lorek}, \citenamefont {Dragan},\ and\ \citenamefont {Omar}}]{Richter2017}%
  \BibitemOpen
  \bibfield  {author} {\bibinfo {author} {\bibfnamefont {B.}~\bibnamefont {Richter}}, \bibinfo {author} {\bibfnamefont {K.}~\bibnamefont {Lorek}}, \bibinfo {author} {\bibfnamefont {A.}~\bibnamefont {Dragan}},\ and\ \bibinfo {author} {\bibfnamefont {Y.}~\bibnamefont {Omar}},\ }\bibfield  {title} {\bibinfo {title} {Effect of acceleration on localized fermionic {G}aussian states: From vacuum entanglement to maximally entangled states},\ }\href {https://doi.org/10.1103/PhysRevD.95.076004} {\bibfield  {journal} {\bibinfo  {journal} {Phys. Rev. D}\ }\textbf {\bibinfo {volume} {95}},\ \bibinfo {pages} {076004} (\bibinfo {year} {2017})}\BibitemShut {NoStop}%
\bibitem [{\citenamefont {Doukas}\ \emph {et~al.}(2013)\citenamefont {Doukas}, \citenamefont {Brown}, \citenamefont {Dragan},\ and\ \citenamefont {Mann}}]{Doukas2013}%
  \BibitemOpen
  \bibfield  {author} {\bibinfo {author} {\bibfnamefont {J.}~\bibnamefont {Doukas}}, \bibinfo {author} {\bibfnamefont {E.~G.}\ \bibnamefont {Brown}}, \bibinfo {author} {\bibfnamefont {A.}~\bibnamefont {Dragan}},\ and\ \bibinfo {author} {\bibfnamefont {R.~B.}\ \bibnamefont {Mann}},\ }\bibfield  {title} {\bibinfo {title} {Entanglement and discord: {A}ccelerated observations of local and global modes},\ }\href {https://doi.org/10.1103/PhysRevA.87.012306} {\bibfield  {journal} {\bibinfo  {journal} {Phys. Rev. A}\ }\textbf {\bibinfo {volume} {87}},\ \bibinfo {pages} {012306} (\bibinfo {year} {2013})}\BibitemShut {NoStop}%
\bibitem [{\citenamefont {Paley}\ and\ \citenamefont {Weiner}(1934)}]{Paley_Wiener}%
  \BibitemOpen
  \bibfield  {author} {\bibinfo {author} {\bibfnamefont {R.~E. A.~C.}\ \bibnamefont {Paley}}\ and\ \bibinfo {author} {\bibfnamefont {N.}~\bibnamefont {Weiner}},\ }\href@noop {} {\emph {\bibinfo {title} {Fourier Transforms in the Complex Domain}}}\ (\bibinfo  {publisher} {American Mathematical Society, New York},\ \bibinfo {year} {1934})\BibitemShut {NoStop}%
\bibitem [{\citenamefont {Reeh}\ and\ \citenamefont {Schlieder}(1961)}]{Reeh1961}%
  \BibitemOpen
  \bibfield  {author} {\bibinfo {author} {\bibfnamefont {H.}~\bibnamefont {Reeh}}\ and\ \bibinfo {author} {\bibfnamefont {S.}~\bibnamefont {Schlieder}},\ }\bibfield  {title} {\bibinfo {title} {Bemerkungen zur {U}nit{\"a}r{\"a}quivalenz von lorentzinvarianten {F}eldern},\ }\href {https://doi.org/10.1007/BF02787889} {\bibfield  {journal} {\bibinfo  {journal} {Nuovo Cimento}\ }\textbf {\bibinfo {volume} {22}},\ \bibinfo {pages} {1051} (\bibinfo {year} {1961})}\BibitemShut {NoStop}%
\bibitem [{\citenamefont {Halvorson}(2001)}]{Halvorson2001}%
  \BibitemOpen
  \bibfield  {author} {\bibinfo {author} {\bibfnamefont {H.}~\bibnamefont {Halvorson}},\ }\bibfield  {title} {\bibinfo {title} {Reeh-{S}chlieder defeats {N}ewton-{W}igner: {O}n alternative localization schemes in relativistic quantum field theory},\ }\href {http://www.jstor.org/stable/3081027} {\bibfield  {journal} {\bibinfo  {journal} {Philos. Sci.}\ }\textbf {\bibinfo {volume} {68}},\ \bibinfo {pages} {111} (\bibinfo {year} {2001})}\BibitemShut {NoStop}%
\bibitem [{\citenamefont {Hegerfeldt}(1974)}]{Hegerfeldt1974}%
  \BibitemOpen
  \bibfield  {author} {\bibinfo {author} {\bibfnamefont {G.~C.}\ \bibnamefont {Hegerfeldt}},\ }\bibfield  {title} {\bibinfo {title} {Remark on causality and particle localization},\ }\href {https://doi.org/10.1103/PhysRevD.10.3320} {\bibfield  {journal} {\bibinfo  {journal} {Phys. Rev. D}\ }\textbf {\bibinfo {volume} {10}},\ \bibinfo {pages} {3320} (\bibinfo {year} {1974})}\BibitemShut {NoStop}%
\bibitem [{\citenamefont {Ferraro}\ \emph {et~al.}(2005)\citenamefont {Ferraro}, \citenamefont {Olivares},\ and\ \citenamefont {Paris}}]{Ferraro}%
  \BibitemOpen
  \bibfield  {author} {\bibinfo {author} {\bibfnamefont {A.}~\bibnamefont {Ferraro}}, \bibinfo {author} {\bibfnamefont {S.}~\bibnamefont {Olivares}},\ and\ \bibinfo {author} {\bibfnamefont {M.}~\bibnamefont {Paris}},\ }\href@noop {} {{\selectlanguage {English}\emph {\bibinfo {title} {Gaussian States in Quantum Information}}}},\ Napoli Series on Physics and Astrophysics\ (\bibinfo  {publisher} {Bibliopolis},\ \bibinfo {address} {Naples},\ \bibinfo {year} {2005})\BibitemShut {NoStop}%
\bibitem [{\citenamefont {Unruh}(1976)}]{Unruh1976}%
  \BibitemOpen
  \bibfield  {author} {\bibinfo {author} {\bibfnamefont {W.~G.}\ \bibnamefont {Unruh}},\ }\bibfield  {title} {\bibinfo {title} {Notes on black-hole evaporation},\ }\href {https://doi.org/10.1103/PhysRevD.14.870} {\bibfield  {journal} {\bibinfo  {journal} {Phys. Rev. D}\ }\textbf {\bibinfo {volume} {14}},\ \bibinfo {pages} {870} (\bibinfo {year} {1976})}\BibitemShut {NoStop}%
\bibitem [{\citenamefont {Nielsen}\ and\ \citenamefont {Chuang}(2000)}]{NielsenChuang2010}%
  \BibitemOpen
  \bibfield  {author} {\bibinfo {author} {\bibfnamefont {M.~A.}\ \bibnamefont {Nielsen}}\ and\ \bibinfo {author} {\bibfnamefont {I.~L.}\ \bibnamefont {Chuang}},\ }\href@noop {} {\emph {\bibinfo {title} {Quantum Computation and Quantum Information}}}\ (\bibinfo  {publisher} {Cambridge University Press},\ \bibinfo {address} {Cambridge},\ \bibinfo {year} {2000})\BibitemShut {NoStop}%
\bibitem [{\citenamefont {Bergou}\ \emph {et~al.}(2004)\citenamefont {Bergou}, \citenamefont {Herzog},\ and\ \citenamefont {Hillery}}]{Bergou2004}%
  \BibitemOpen
  \bibfield  {author} {\bibinfo {author} {\bibfnamefont {J.~A.}\ \bibnamefont {Bergou}}, \bibinfo {author} {\bibfnamefont {U.}~\bibnamefont {Herzog}},\ and\ \bibinfo {author} {\bibfnamefont {M.}~\bibnamefont {Hillery}},\ }\bibinfo {title} {Discrimination of quantum states},\ in\ \href {https://doi.org/10.1007/978-3-540-44481-7_11} {\emph {\bibinfo {booktitle} {Quantum State Estimation}}},\ \bibinfo {editor} {edited by\ \bibinfo {editor} {\bibfnamefont {M.}~\bibnamefont {Paris}}\ and\ \bibinfo {editor} {\bibfnamefont {J.}~\bibnamefont {{\v{R}}eh{\'a}{\v{c}}ek}}}\ (\bibinfo  {publisher} {Springer Berlin Heidelberg},\ \bibinfo {address} {Berlin, Heidelberg},\ \bibinfo {year} {2004})\ pp.\ \bibinfo {pages} {417--465}\BibitemShut {NoStop}%
\bibitem [{\citenamefont {Helstrom}(1976)}]{Helstrom1976}%
  \BibitemOpen
  \bibfield  {author} {\bibinfo {author} {\bibfnamefont {C.~W.}\ \bibnamefont {Helstrom}},\ }\href@noop {} {\emph {\bibinfo {title} {{Quantum Detection and Estimation Theory}}}}\ (\bibinfo  {publisher} {Academic Press, New York},\ \bibinfo {year} {1976})\BibitemShut {NoStop}%
\bibitem [{\citenamefont {Weedbrook}\ \emph {et~al.}(2012)\citenamefont {Weedbrook}, \citenamefont {Pirandola}, \citenamefont {Garc\'{\i}a-Patr\'on}, \citenamefont {Cerf}, \citenamefont {Ralph}, \citenamefont {Shapiro},\ and\ \citenamefont {Lloyd}}]{Weedbrook2012}%
  \BibitemOpen
  \bibfield  {author} {\bibinfo {author} {\bibfnamefont {C.}~\bibnamefont {Weedbrook}}, \bibinfo {author} {\bibfnamefont {S.}~\bibnamefont {Pirandola}}, \bibinfo {author} {\bibfnamefont {R.}~\bibnamefont {Garc\'{\i}a-Patr\'on}}, \bibinfo {author} {\bibfnamefont {N.~J.}\ \bibnamefont {Cerf}}, \bibinfo {author} {\bibfnamefont {T.~C.}\ \bibnamefont {Ralph}}, \bibinfo {author} {\bibfnamefont {J.~H.}\ \bibnamefont {Shapiro}},\ and\ \bibinfo {author} {\bibfnamefont {S.}~\bibnamefont {Lloyd}},\ }\bibfield  {title} {\bibinfo {title} {Gaussian quantum information},\ }\href {https://doi.org/10.1103/RevModPhys.84.621} {\bibfield  {journal} {\bibinfo  {journal} {Rev. Mod. Phys.}\ }\textbf {\bibinfo {volume} {84}},\ \bibinfo {pages} {621} (\bibinfo {year} {2012})}\BibitemShut {NoStop}%
\bibitem [{\citenamefont {Scutaru}(1998)}]{Scutaru1998}%
  \BibitemOpen
  \bibfield  {author} {\bibinfo {author} {\bibfnamefont {H.}~\bibnamefont {Scutaru}},\ }\bibfield  {title} {\bibinfo {title} {Fidelity for displaced squeezed thermal states and the oscillator semigroup},\ }\href {https://doi.org/10.1088/0305-4470/31/15/025} {\bibfield  {journal} {\bibinfo  {journal} {J. Phys. A}\ }\textbf {\bibinfo {volume} {31}},\ \bibinfo {pages} {3659} (\bibinfo {year} {1998})}\BibitemShut {NoStop}%
\bibitem [{\citenamefont {Nha}\ and\ \citenamefont {Carmichael}(2005)}]{Nha2005}%
  \BibitemOpen
  \bibfield  {author} {\bibinfo {author} {\bibfnamefont {H.}~\bibnamefont {Nha}}\ and\ \bibinfo {author} {\bibfnamefont {H.~J.}\ \bibnamefont {Carmichael}},\ }\bibfield  {title} {\bibinfo {title} {Distinguishing two single-mode {G}aussian states by homodyne detection: {A}n information-theoretic approach},\ }\href {https://doi.org/10.1103/PhysRevA.71.032336} {\bibfield  {journal} {\bibinfo  {journal} {Phys. Rev. A}\ }\textbf {\bibinfo {volume} {71}},\ \bibinfo {pages} {032336} (\bibinfo {year} {2005})}\BibitemShut {NoStop}%
\bibitem [{\citenamefont {Olivares}\ \emph {et~al.}(2006)\citenamefont {Olivares}, \citenamefont {Paris},\ and\ \citenamefont {Andersen}}]{Olivares2006}%
  \BibitemOpen
  \bibfield  {author} {\bibinfo {author} {\bibfnamefont {S.}~\bibnamefont {Olivares}}, \bibinfo {author} {\bibfnamefont {M.~G.~A.}\ \bibnamefont {Paris}},\ and\ \bibinfo {author} {\bibfnamefont {U.~L.}\ \bibnamefont {Andersen}},\ }\bibfield  {title} {\bibinfo {title} {Cloning of {G}aussian states by linear optics},\ }\href {https://doi.org/10.1103/PhysRevA.73.062330} {\bibfield  {journal} {\bibinfo  {journal} {Phys. Rev. A}\ }\textbf {\bibinfo {volume} {73}},\ \bibinfo {pages} {062330} (\bibinfo {year} {2006})}\BibitemShut {NoStop}%
\bibitem [{\citenamefont {Holevo}(1975)}]{Holevo1975}%
  \BibitemOpen
  \bibfield  {author} {\bibinfo {author} {\bibfnamefont {A.}~\bibnamefont {Holevo}},\ }\bibfield  {title} {\bibinfo {title} {Some statistical problems for quantum {G}aussian states},\ }\href {https://doi.org/10.1109/TIT.1975.1055441} {\bibfield  {journal} {\bibinfo  {journal} {IEEE Trans. Inf. Theory}\ }\textbf {\bibinfo {volume} {21}},\ \bibinfo {pages} {533} (\bibinfo {year} {1975})}\BibitemShut {NoStop}%
\bibitem [{\citenamefont {Fuchs}\ and\ \citenamefont {van~de Graaf}(1999)}]{Fuchs_Graff1999}%
  \BibitemOpen
  \bibfield  {author} {\bibinfo {author} {\bibfnamefont {C.}~\bibnamefont {Fuchs}}\ and\ \bibinfo {author} {\bibfnamefont {J.}~\bibnamefont {van~de Graaf}},\ }\bibfield  {title} {\bibinfo {title} {Cryptographic distinguishability measures for quantum-mechanical states},\ }\href {https://doi.org/10.1109/18.761271} {\bibfield  {journal} {\bibinfo  {journal} {IEEE Trans. Inf. Theory}\ }\textbf {\bibinfo {volume} {45}},\ \bibinfo {pages} {1216} (\bibinfo {year} {1999})}\BibitemShut {NoStop}%
\bibitem [{\citenamefont {Crispino}\ \emph {et~al.}(2008)\citenamefont {Crispino}, \citenamefont {Higuchi},\ and\ \citenamefont {Matsas}}]{Crispino2008}%
  \BibitemOpen
  \bibfield  {author} {\bibinfo {author} {\bibfnamefont {L.~C.~B.}\ \bibnamefont {Crispino}}, \bibinfo {author} {\bibfnamefont {A.}~\bibnamefont {Higuchi}},\ and\ \bibinfo {author} {\bibfnamefont {G.~E.~A.}\ \bibnamefont {Matsas}},\ }\bibfield  {title} {\bibinfo {title} {The {U}nruh effect and its applications},\ }\href {https://doi.org/10.1103/RevModPhys.80.787} {\bibfield  {journal} {\bibinfo  {journal} {Rev. Mod. Phys.}\ }\textbf {\bibinfo {volume} {80}},\ \bibinfo {pages} {787} (\bibinfo {year} {2008})}\BibitemShut {NoStop}%
\bibitem [{\citenamefont {Ch\ifmmode \mbox{\k{e}}\else \k{e}\fi{}ci\ifmmode~\acute{n}\else \'{n}\fi{}ska}\ and\ \citenamefont {Dragan}(2015)}]{Checinska2015}%
  \BibitemOpen
  \bibfield  {author} {\bibinfo {author} {\bibfnamefont {A.}~\bibnamefont {Ch\ifmmode \mbox{\k{e}}\else \k{e}\fi{}ci\ifmmode~\acute{n}\else \'{n}\fi{}ska}}\ and\ \bibinfo {author} {\bibfnamefont {A.}~\bibnamefont {Dragan}},\ }\bibfield  {title} {\bibinfo {title} {Communication between general-relativistic observers without a shared reference frame},\ }\href {https://doi.org/10.1103/PhysRevA.92.012321} {\bibfield  {journal} {\bibinfo  {journal} {Phys. Rev. A}\ }\textbf {\bibinfo {volume} {92}},\ \bibinfo {pages} {012321} (\bibinfo {year} {2015})}\BibitemShut {NoStop}%
\bibitem [{\citenamefont {Falcone}\ and\ \citenamefont {Conti}(2023)}]{Falcone2023}%
  \BibitemOpen
  \bibfield  {author} {\bibinfo {author} {\bibfnamefont {R.}~\bibnamefont {Falcone}}\ and\ \bibinfo {author} {\bibfnamefont {C.}~\bibnamefont {Conti}},\ }\bibfield  {title} {\bibinfo {title} {Observing single particles beyond the {R}indler horizon},\ }\href {https://doi.org/10.1103/PhysRevA.107.L030203} {\bibfield  {journal} {\bibinfo  {journal} {Phys. Rev. A}\ }\textbf {\bibinfo {volume} {107}},\ \bibinfo {pages} {L030203} (\bibinfo {year} {2023})}\BibitemShut {NoStop}%
\bibitem [{\citenamefont {Michalski}(2026)}]{Michalski2026}%
  \BibitemOpen
  \bibfield  {author} {\bibinfo {author} {\bibfnamefont {P.}~\bibnamefont {Michalski}},\ }\bibfield  {title} {\bibinfo {title} {Detection of quantum entanglement across the event horizon},\ }\bibfield  {journal} {\bibinfo  {journal} {Zenodo}\ }\href {https://doi.org/10.5281/zenodo.20801173} {10.5281/zenodo.20801173} (\bibinfo {year} {2026})\BibitemShut {NoStop}%
\end{thebibliography}%

\end{document}